# Coherent diffraction and holographic imaging of individual biomolecules using low-energy electrons

Tatiana Latychevskaia*, Jean-Nicolas Longchamp, Conrad Escher, Hans-Werner Fink

Physics Institute, University of Zurich
Winterthurerstrasse 190, 8057 Zurich,
Switzerland

*Email: tatiana@physik.uzh.ch

**Abstract** Modern microscopy techniques are aimed at imaging an individual molecule at atomic resolution. Here we show that low-energy electrons with kinetic energies of 50-250 eV offer a possibility of overcome the problem of radiation damage, and obtaining images of individual biomolecules. Two experimental schemes for obtaining images of individual molecules – holography and coherent diffraction imaging – are discussed and compared. Images of individual molecules obtained by both techniques, using low-energy electrons, are shown.

## Introduction

Investigating the structure of biomolecules at the atomic scale has always been of utmost importance for healthcare, medicine and life science in general, since the three-dimensional shape of proteins, for example, relates to their function. At the moment, these structural data are predominantly obtained by X-ray crystallography, cryo-electron microscopy and NMR. Despite there being an impressive database (www.pdb.org) obtained with these methods, they all require large quantities of a particular protein. This leads to averaging over fine conformational details in the recovered structure. The goal of modern imaging techniques is to visualize *an individual biomolecule* at atomic resolution.

## Imaging an individual molecule: choice of radiation

A direct visualization of an individual molecule at Ångstrom resolution can be achieved using electron or X-ray waves which have a wavelength of about 1 Å, see Fig. 1. Although both, X-rays and high-energy electrons possess sufficiently short wavelengths to resolve the individual atoms constituting a protein, the resolution achieved is mainly limited by radiation damage inherent to both types of radiation.

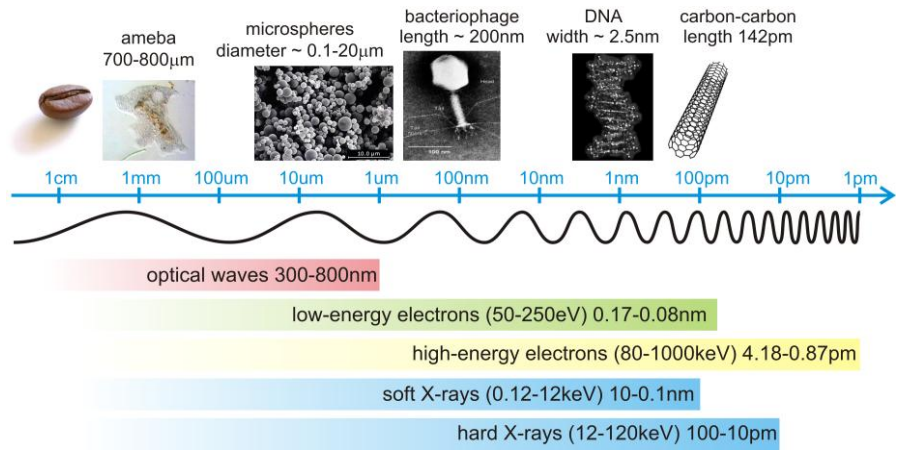

**Fig. 1.** Spectrum of radiation used for imaging. The bars show the range of the sizes of the objects which can be imaged with the assigned radiation.

### *Imaging with high-energy electrons (80-200 keV)*

In cryo-electron microscopy (Adrian et al. 1984), cooling the sample to the temperature of liquid nitrogen allows a higher electron dose to be used for the same amount of radiation damage. Depending on the resolution required, typical electron exposures vary between 5 and 25 e/Å$^2$ (Henderson 2004). Due to the very low signal-to-noise ratio in the images obtained, over 10,000 images of individual molecules typically need to be collected and averaged to arrive at the reconstruction of the structure (van Heel et al. 2000).

### *Imaging with X-rays*

Visualization of an individual molecule at atomic resolution by employing X-rays is planned at the X-ray Free Electron Lasers (XFELs) facilities which are being developed worldwide. Here the radiation damage problem (Howells et al. 2009) is circumvented by employing ultra-short X-ray pulses, which allow the diffraction pattern of an individual molecule to be recorded before it decomposes due to the strong inelastic scattering (Neutze et al. 2000; Bergh et al. 2008). High intensity X-ray laser pulses will provide the intensity in the diffraction pattern detected at the high scattering angles, which is required for further numerical recovery of the molecular structure at high resolution.

### *Imaging with low-energy electrons (50-250 eV)*

Low-energy electrons (with kinetic energies of 50-250 eV, corresponding to wavelengths of 0.78-1.73 Å) can be employed to visualize individual biomolecules directly. It has been shown (Germann et al. 2010) that individual DNA molecules can withstand low-energy electron radiation having energy of 60 eV (corresponding to a wavelength of 1.58 Å) for at least 70 minutes. This in total amounts to a radiation dose of 10$^6$ e/Å$^2$, which is at least six orders of magnitude larger than the permissible dose in high-energy electron microscopy or X-ray imaging.

## Imaging an individual molecule: the phase problem

The principle of lensless imaging of an individual molecule is as follows: when a coherent wave is scattered by a molecule, it carries both, amplitude and phase information imposed by the scattering event. The phase distribution is especially important since it carries information about the position of the atoms constituting the molecule. However, detectors are not sensitive to the phase information; instead they just record the intensity which is the square of the wave amplitude. Hence, the recovery of the complex-valued scattered wave requires a solution to the so-called *phase problem*. Today there are two known solutions to the phase retrieval problem: holography and coherent diffraction imaging (CDI), both schematically shown in Fig. 2. Their proper implementation would ultimately allow the atomic mapping of an individual molecule in three dimensions.

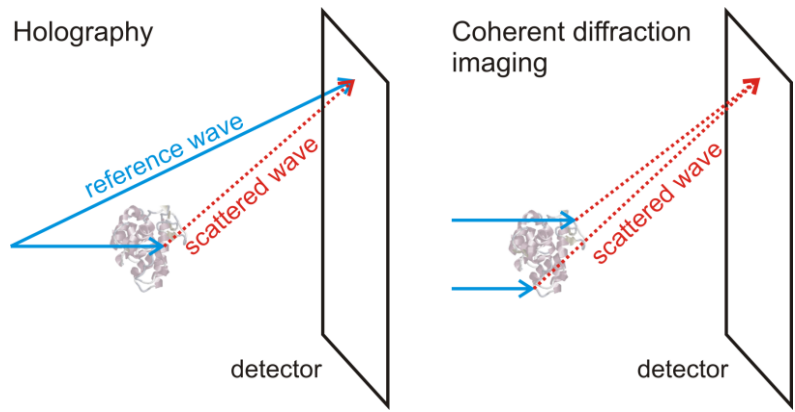

**Fig. 2.** Schematics of the lensless imaging of an individual molecule.

In Fig. 3, the experimental set-ups for both, holographic and CDI recording with low-energy electrons designed and built in our laboratory (Fink et al. 1990; Steinwand et al. 2011) are sketched.

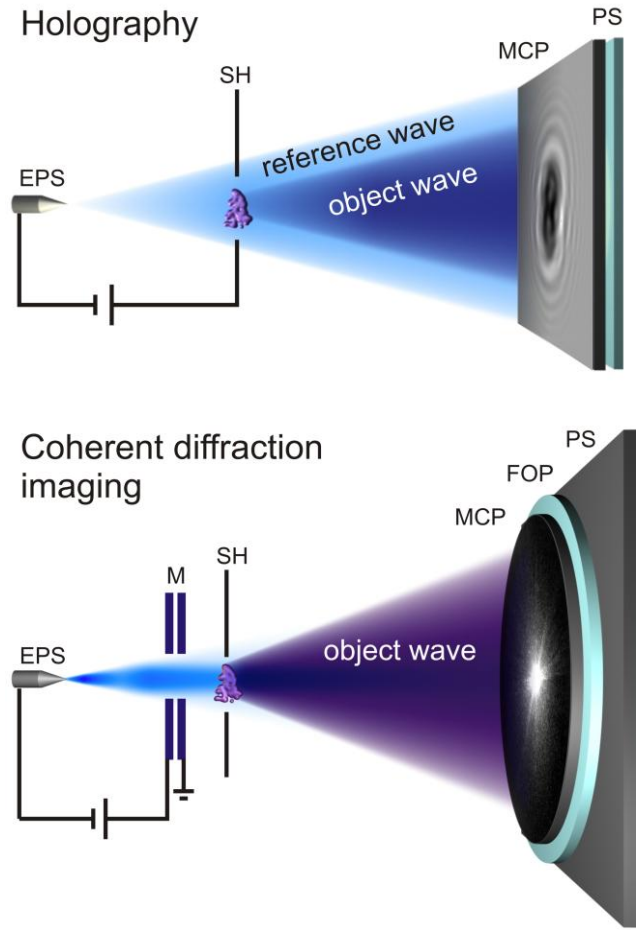

**Fig. 3.** Low-energy electron microscopes. Coherent low-energy electrons are extracted from the electron point source (EPS) by field emission. A biological molecule is fixed in the sample holder (SH). In the holographic microscope, the interference between the scattered (object) wave and unscattered (reference) wave is recorded by the detector unit (consisting of a micro-channel plate (MCP) followed by a phosphor screen (PS)). In the CDI microscope, the electron beam is collimated by a microlens (ML) and the detector unit consists of an MCP followed by a fibre optic plate (FOP) with a thin phosphorous layer (PS).

## Holography

In holography the unknown wave that is scattered by an object is superimposed with a known reference wave. A hologram is the interference pattern formed by constructive and destructive interference between these two waves (Gabor 1949) and is illustrated in Fig. 4. The holography technique uniquely solves the phase problem in one step because of the presence of the reference wave. However, it lacks high resolution due to the higher-order scattered signal being buried in the experimental noise of the reference wave.

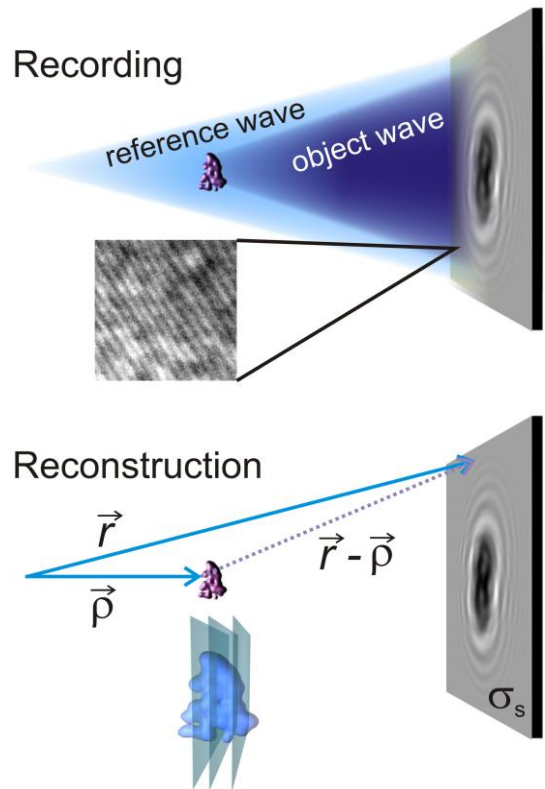

**Fig. 4.** Illustration of recording an inline hologram and reconstructing the object. Recording: the superposition of the reference wave and the wave scattered by the object is recorded; a magnified region of the hologram shows the fringes of the interference pattern. Reconstruction: back propagation of the recovered object wave from the hologram plane to the planes of the object's location (analogous to optical sectioning) results in a three-dimensional reconstruction.

Hologram reconstruction includes two steps: (1) illumination of the hologram with the reference wave and (2) backward propagation of the wavefront to the position of the object. In numerical reconstruction, the complex-valued reference wave at the hologram plane is simulated and the propagation from the hologram back to the object is calculated using Huygens' principle and Fresnel formalism:

$$U(\vec{\rho}) = \frac{i}{\lambda} \iint H(\vec{r}) \frac{\exp(ikr)}{r} \frac{\exp(-ik|\vec{r}-\vec{\rho}|)}{|\vec{r}-\vec{\rho}|} d\sigma_s, \qquad (1)$$

where $H(\mathbf{r})$ is the hologram's transmission function distribution, $\mathbf{r}$ and $\mathbf{\rho}$ are defined as illustrated in Fig. 4, and the integration is performed over the hologram's surface $\sigma_s$. The result of this integral transform is a complex-valued distribution of the object wavefront at any coordinate $\rho$, and, hence, a three-dimensional reconstruction.

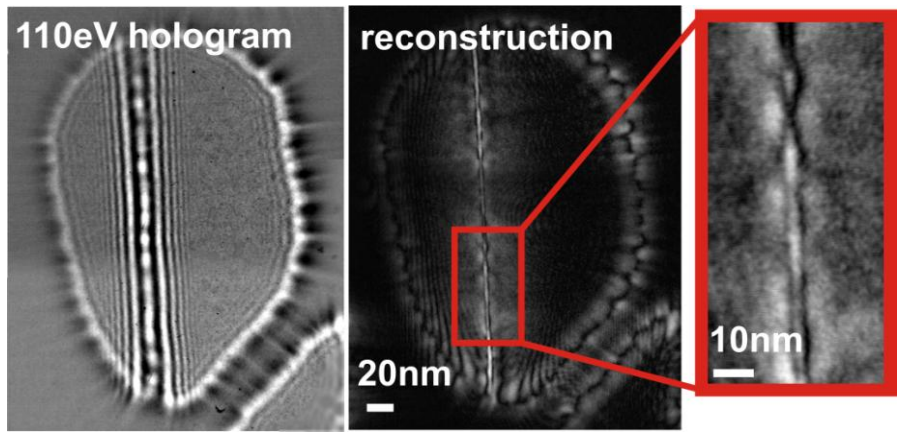

**Fig. 5.** Low-energy electron hologram of an individual ssDNA molecule stretched over a hole in a carbon film (sample courtesy by Michael and William Andregg, www.halcyonmolecular.com).

An example of an inline hologram of an individual DNA molecule and its reconstruction is shown in Fig. 5. The successful trials during the last decade of imaging individual biological molecules by low-energy electron holography include the imaging of: DNA molecules (Fink et al. 1997; Eisele et al. 2008), phthalocyaninato polysiloxane molecules (Golzhauser et al. 1998), the tobacco mosaic virus (Weierstall et al. 1999), a bacteriophage (Stevens et al. 2011) and ferritin (Longchamp et al. 2012). Despite a very short wavelength (1-2 Å) of the probing electron wave, the resolution in the reconstructed molecular structures remains in the order of a few nanometres. The reason is that the resolution in inline holography is limited by the detectability of the interference fringes at high diffraction angles (Spence et al. 1994; Latychevskaia et al. 2011) (such as, for instance, the fringes shown in the magnified region in Fig. 4). The pattern of these fine fringes is very sensitive to the object's lateral movements and can be destroyed by the object shifting even by just the distance corresponding to the wavelength. In addition, these fine fringes are often buried in the experimental noise of the reference wave.

**Coherent diffraction imaging**

CDI is a relatively modern technique which combines the recording of a far-field diffraction pattern of a non-crystalline object and the numerical recovery of the object structure. In 1952, Sayre proposed that it was possible to recover the phase information associated with scattering off a non-crystalline specimen by sampling its diffraction pattern at a frequency higher than twice the Nyquist frequency (oversampling) (Sayre 1952). In 1972, Gerchberg and Saxton proposed an iterative algorithm to recover the phase distribution from two amplitude measurements taken: at the object plane and at the far-field plane (Gerchberg and Saxton 1972). In 1998, Miao et al. combined these two ideas and successfully recovered an object from its oversampled diffraction pattern (Miao et al. 1998). They demonstrated that the phase retrieval algorithm converges if the initial conditions are such that the surrounding of the molecule ("support") is known. The concept of knowing the support of the molecule is analogous to the solvent flattening technique in the phasing methods. The known surrounding of a molecule is usually mathematically described by zero-padding the object, which in turn leads to oversampling of the spectrum in the Fou-rier domain. Thus, reconstruction becomes possible if the diffraction pattern is recorded under the oversampling condition (Miao et al. 1999; Miao and Sayre 2000; Miao et al. 2003b); this is also illustrated in Fig. 6.

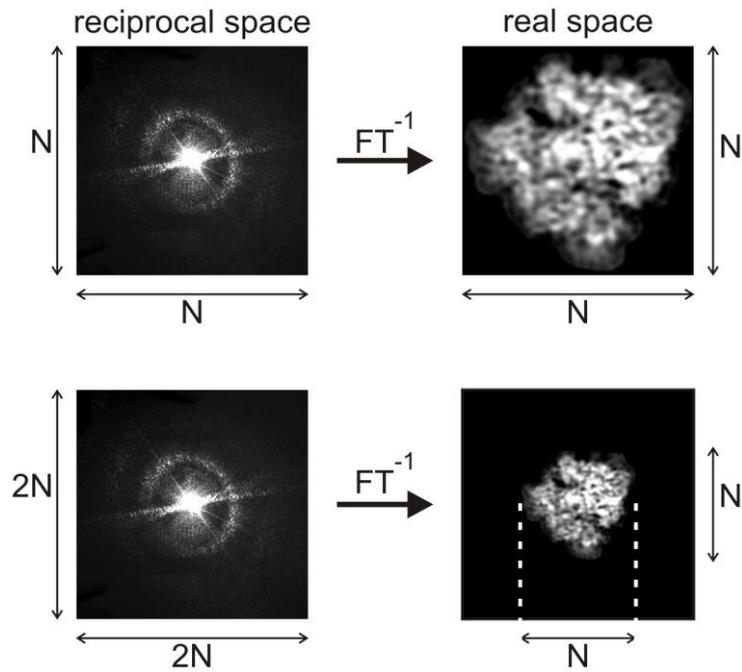

**Fig. 6.** Sampling at the Nyquist frequency (upper row) and twice the Nyquist frequency (lower row). The Fourier transform of the spectrum sampled at the Nyquist frequency results in the object distribution filling the entire reconstructed area. The Fourier transform of the spectrum sampled at twice the Nyquist frequency results in the zero-padded object distribution (Miao and Sayre 2000).

The basic iterative reconstruction loop (Fienup 1982) is shown in Fig. 7. It begins with the complex-valued wave distribution at the detector plane which is formed by the superposition of the square root of the measured intensity and a random phase distribution. In the object domain various constraints are applied. For instance, the electron density reconstructed from the X-ray diffraction images must be real and positive.

The resolution in CDI is defined by the outermost detected signal in the diffraction pattern, $R=\lambda/\sin\theta$, where $\theta$ is the scattering angle. The resolving power of the CDI technique has already been demonstrated by the reconstruction of a double-walled carbon nanotube at a resolution of 1 Å from a coherent diffraction pattern recorded using a 200 keV electron microscope exhibiting a nominal conventional TEM resolution of 2.2 Å (Zuo et al. 2003).

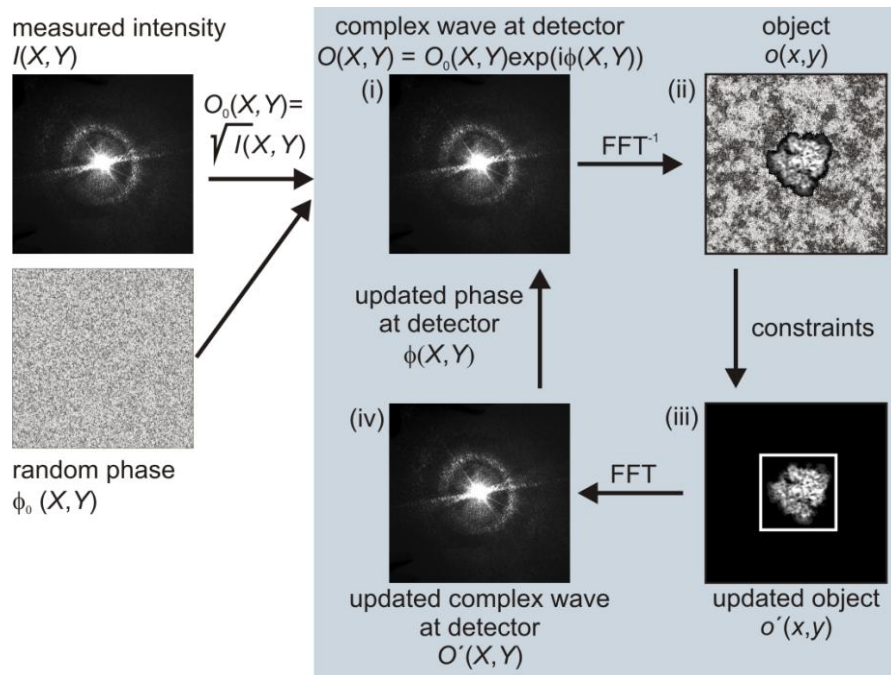

**Fig. 7.** Iterative reconstruction of a coherent diffraction pattern. Left column: amplitude and phase distributions at the detector plane initiating the iterative loop. Right: the steps (i)-(iv) showing the flow of the iterative loop.

The X-ray diffraction pattern of a crystal, unlike that of an individual molecule, displays a strong signal due to the periodicity of the crystal. Obtaining an X-ray diffraction pattern of *an individual molecule* in turn requires a much more intense X-ray beam. As a consequence, the resolution is limited by radiation damage and remains very moderate. A few biological specimens have been imaged by CDI using X-rays at a resolution of a few nanometres: E.coli bacteria (Miao et al. 2003a), an unstained yeast cell (Shapiro et al. 2005), single herpes virions (Song et al. 2008), malaria-infected red blood cells (Williams et al. 2008), a frozen hydrated yeast cell (Huang et al. 2009), human chromosomes (Nishino et al. 2009), unstained and unsliced freeze-dried cells of the bacterium Deinococcus radiodurans by ptychography (Giewekemeyer et al. 2010), and labelled yeast cells (Nelson et al. 2010). Ultra-short and extremely bright coherent X-ray pulses from XFEL allow the recording of a high-resolution diffraction pattern before the sample explodes (Neutze et al. 2000; Bergh et al. 2008). The first results from the first XFEL facility to be operational (the Linac Coherent Light Source) reported imaging an individual unstained mimivirus at 32 nanometre resolution (Seibert et al. 2011); in this experiment an X-ray pulse of 1.8 keV (6.9Å) energy and 70 fs duration was focused to a spot 10μm in diameter with $1.6 \times 10^{10}$ photons per 1 μm$^2$. A sub-nanometre resolution could be achieved by employing shorter pulses and a higher photon flux (Bergh et al. 2008; Seibert et al. 2011); at present this is beyond the capabilities of the XFELs but might be realized with the next generation of XFELs.

**Comparing holography and CDI**

Each of the two techniques has its pros (+) and cons (-), which are summarized below:

*Holography*

• Requires well-defined reference wave over entire detector area (-)
• Non-iterative reconstruction by calculating back-propagation integral (+)
• Three-dimensional reconstruction (+)
• Low resolution, due to high sensitivity of the interference pattern to object shifts and experimental noise in the reference wave (-)

*CDI*

- No reference wave is required (+)
- Reconstruction is done by an iterative procedure and does not always converge to a uniquely defined outcome (-)
- Reconstruction is not three-dimensional, it is limited to one plane (-)
- High resolution provided by stability of diffraction pattern being insensitive to shifts of the object (+)

## HCDI: Combining holography and CDI

Recently, we revealed the relationship between the hologram and the diffraction pattern of an object, which allows holography and CDI to be combined into a superior technique: holographic coherent diffraction imaging (HCDI). HCDI inherits fast and reliable reconstruction from holography and the highest possible resolution from CDI(Latychevskaia et al. 2012).

The Fourier transform of an inline hologram is proportional to the complex-valued object wave in the far-field, as is illustrated for experimental images in Fig. 8. Thus, the phase distribution of the Fourier transform of the inline hologram provides the phase distribution of the object wave in the far-field and hence the solution to the "phase problem" in just one step. The diffraction pattern is then required to refine the reconstruction of the high-resolution information by a conventional iterative procedure. In addition, the central region of the diffraction pattern, which is usually missing, can be adapted from the amplitude of the Fourier transform of the hologram; see Fig. 8d.

The hologram and the diffraction pattern of a bundle of carbon nanotubes recorded with the coherent low-energy electron diffraction microscope (Steinwand et al. 2011) are shown in Fig. 9. The HCDI technique was applied to reconstruct these images and the result is shown in Fig. 9d.

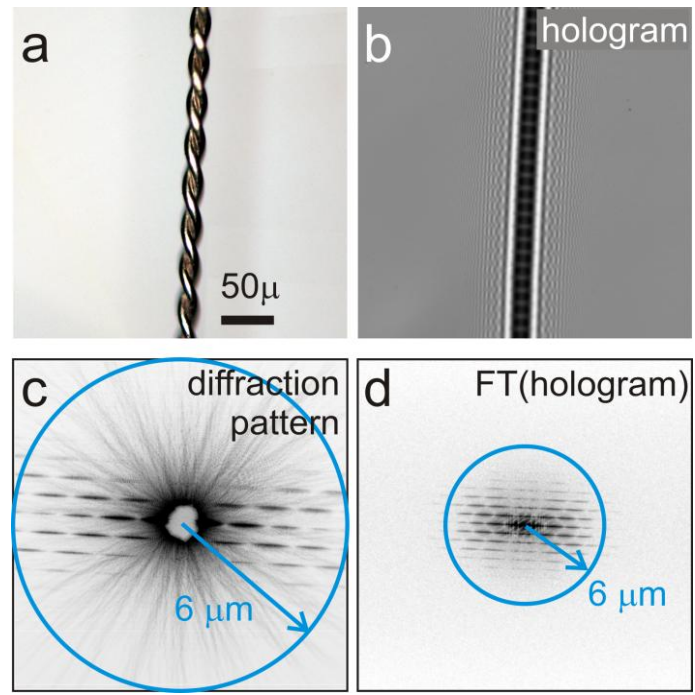

**Fig. 8.** Experimental verification of the relationship between a hologram and a diffraction pattern. (a) Reflected-light microscopy image of two twisted tungsten wires. (b) Inline hologram recorded with laser light. (c) Diffraction pattern. (d) The amplitude of the Fourier transform of the hologram is displayed using a logarithmic and inverted intensity scale. The diffraction pattern provides the same resolution as the hologram - namely 6 μm – but it is recorded while fulfilling the oversampling condition (Latychevskaia et al. 2012).

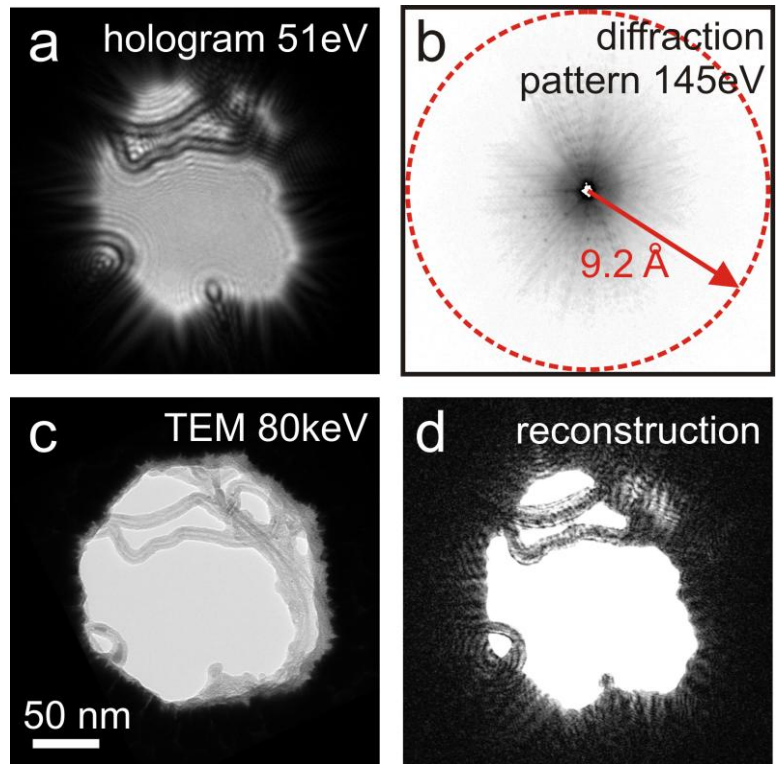

**Fig. 9.** (a) Hologram of carbon nanotubes acquired with 51 eV electrons. (b) Coherent diffraction pattern of the same nanotubes recorded with electrons of 145 eV kinetic energy. The dashed circle shows the highest order signal detected corresponding to 9.2 Å resolution. (c) TEM image of the very same area obtained with 80 keV electrons. (d) Reconstruction obtained by the HCDI method (Latychevskaia et al. 2012).

Because the phase distribution stored in a holographic image is uniquely defined and is associated with the three-dimensional object distribution, HCDI may offer the possibility of retrieving a three-dimensional object distribution from its diffraction pattern.

**Acknowledgments** We would like to thank the Swiss National Science Foundation for its financial support.

16Miao JW, Sayre D (2000) On possible extensions of x-ray crystallography through diffraction-pattern oversampling. Acta Crystallogr A 56:596-605

Miao JW, Sayre D, Chapman HN (1998) Phase retrieval from the magnitude of the Fourier transforms of nonperiodic objects. Journal of the Optical Society of America A - Optics Image Science and Vision 15 (6):1662-1669

Nelson J, Huang XJ, Steinbrener J, Shapiro D, Kirz J, Marchesini S, Neiman AM, Turner JJ, Jacobsen C (2010) High-resolution x-ray diffraction microscopy of specifically labeled yeast cells. Proc Natl Acad Sci USA 107 (16):7235-7239

Neutze R, Wouts R, van der Spoel D, Weckert E, Hajdu J (2000) Potential for biomolecular imaging with femtosecond x-ray pulses. Nature 406 (6797):752-757

Nishino Y, Takahashi Y, Imamoto N, Ishikawa T, Maeshima K (2009) Three-dimensional visualization of a human chromosome using coherent x-ray diffraction. Phys Rev Lett 102 (1):018101

Sayre D (1952) Some implications of a theorem due to Shannon. Acta Crystallogr 5 (6):843-843

Seibert MM, Ekeberg T, Maia FRNC, Svenda M, Andreasson J, Jonsson O, Odic D, Iwan B, Rocker A, Westphal D, Hantke M, DePonte DP, Barty A, Schulz J, Gumprecht L, Coppola N, Aquila A, Liang M, White TA, Martin A, Caleman C, Stern S, Abergel C, Seltzer V, Claverie J-M, Bostedt C, Bozek JD, Boutet S, Miahnahri AA, Messerschmidt M, Krzywinski J, Williams G, Hodgson KO, Bogan MJ, Hampton CY, Sierra RG, Starodub D, Andersson I, Bajt S, Barthelmess M, Spence JCH, Fromme P, Weierstall U, Kirian R, Hunter M, Doak RB, Marchesini S, Hau-Riege SP, Frank M, Shoeman RL, Lomb L, Epp SW, Hartmann R, Rolles D, Rudenko A, Schmidt C, Foucar L, Kimmel N, Holl P, Rudek B, Erk B, Homke A, Reich C, Pietschner D, Weidenspointner G, Struder L, Hauser G, Gorke H, Ullrich J, Schlichting I, Herrmann S, Schaller G, Schopper F, Soltau H, Kuhnel K-U, Andritschke R, Schroter C-D, Krasniqi F, Bott M, Schorb S, Rupp D, Adolph M, Gorkhover T, Hirsemann H, Potdevin G, Graafsma H, Nilsson B, Chapman HN, Hajdu J (2011) Single mimivirus particles intercepted and imaged with an x-ray laser. Nature 470 (7332):78-81

Shapiro D, Thibault P, Beetz T, Elser V, Howells M, Jacobsen C, Kirz J, Lima E, Miao H, Neiman AM, Sayre D (2005) Biological imaging by soft x-ray diffraction microscopy. Proc Natl Acad Sci USA 102 (43):15343-15346

Song CY, Jiang HD, Mancuso A, Amirbekian B, Peng L, Sun R, Shah SS, Zhou ZH, Ishikawa T, Miao JW (2008) Quantitative imaging of single, unstained viruses with coherent x-rays. Phys Rev Lett 101 (15):158101

Spence JCH, Qian W, Silverman MP (1994) Electron source brightness and degeneracy from Fresnel fringes in-field emission point projection microscopy. J Vac Sci Technol A - Vac Surf Films 12 (2):542-547

Steinwand E, Longchamp J-N, Fink H-W (2011) Coherent low-energy electron diffraction on individual nanometer sized objects. Ultramicroscopy 111 (4):282-284

Stevens GB, Krüger M, Latychevskaia T, Lindner P, Plückthun A, Fink H-W (2011) Individual filamentous phage imaged by electron holography. Eur Biophys J 40:1197-1201 van Heel M, Gowen B, Matadeen R, Orlova EV, Finn R, Pape T, Cohen D, Stark H, Schmidt R, Schatz M, Patwardhan A (2000) Single-particle electron cryo-microscopy: towards atomic resolution. Q Rev Biophys 33 (4):307-369

Weierstall U, Spence JCH, Stevens M, Downing KH (1999) Point-projection electron imaging of tobacco mosaic virus at 40 eV electron energy. Micron 30 (4):335-338

Williams GJ, Hanssen E, Peele AG, Pfeifer MA, Clark J, Abbey B, Cadenazzi G, de Jonge MD, Vogt S, Tilley L, Nugent KA (2008) High-resolution x-ray imaging of plasmodium falciparum-infected red blood cells. Cytom Part A 73A (10):949-957

www.pdb.org Protein Data Bank.